\documentclass[sigconf, nonacm]{acmart}

\AtBeginDocument{%
  }

\setcopyright{none}

\usepackage{multirow}
\usepackage{stfloats}
\usepackage{makecell}
\usepackage{balance}
\usepackage{svg}

\definecolor{lavender}{HTML}{E8DAEF}
\definecolor{mint}{HTML}{D1F2EB}
\definecolor{lightblue}{HTML}{D6EAF8}
\definecolor{green_oh}{HTML}{45A647}
\definecolor{red_oh}{HTML}{F02526}

\newcommand{\blackcircle}{\large$\bullet$}

\begin{document}

\title{MM-HSD: Multi-Modal Hate Speech Detection in Videos}

\author{Berta Céspedes-Sarrias}
\authornote{These authors contributed equally to this research.}
\affiliation{%
  \institution{EPFL}
  \city{Lausanne}
  \country{Switzerland}
}

\affiliation{%
  \institution{Idiap Research Institute}
  \city{Martigny}
  \country{Switzerland}
}

\email{berta.cespedessarrias@epfl.ch}

\author{Carlos Collado-Capell}
\authornotemark[1]

\affiliation{%
  \institution{EPFL}
  \city{Lausanne}
  \country{Switzerland}
}

\affiliation{%
  \institution{Idiap Research Institute}
  \city{Martigny}
  \country{Switzerland}
}

\email{carlos.colladocapell@epfl.ch}

\author{Pablo Rodenas-Ruiz}
\authornotemark[1]

\affiliation{%
  \institution{EPFL}
  \city{Lausanne}
  \country{Switzerland}
}

\affiliation{%
  \institution{Idiap Research Institute}
  \city{Martigny}
  \country{Switzerland}
}

\email{pablo.rodenasruiz@epfl.ch}

\author{Olena Hrynenko}

\affiliation{%
  \institution{EPFL}
  \city{Lausanne}
  \country{Switzerland}
}

\affiliation{%
  \institution{Idiap Research Institute}
  \city{Martigny}
  \country{Switzerland}
}

\email{olena.hrynenko@epfl.ch}

\author{Andrea Cavallaro}

\affiliation{%
  \institution{EPFL}
  \city{Lausanne}
  \country{Switzerland}
}

\affiliation{%
  \institution{Idiap Research Institute}
  \city{Martigny}
  \country{Switzerland}
}

\email{andrea.cavallaro@epfl.ch}

\begin{abstract}
While hate speech detection (HSD) has been extensively studied in text, existing multi-modal approaches remain limited, particularly in videos. As modalities are not always individually informative, simple fusion methods fail to fully capture inter-modal dependencies. Moreover, previous work often omits relevant modalities such as on-screen text and audio, which may contain subtle hateful content and thus provide essential cues, both individually and in combination with others. In this paper, we present MM-HSD, a multi-modal model for HSD in videos that integrates video frames, audio, and text derived from speech transcripts and from frames (i.e.~on-screen text) together with features extracted by Cross-Modal Attention (CMA). We are the first to use CMA as an early feature extractor for HSD in videos, to systematically compare query/key configurations, and to evaluate the interactions between different modalities in the CMA block.
Our approach leads to improved performance when on-screen text is used as a query and the rest of the modalities serve as a key.
Experiments on the HateMM dataset show that MM-HSD outperforms state-of-the-art methods on M-F1 score (0.874), using concatenation of transcript, audio, video, on-screen text, and CMA for feature extraction on raw embeddings of the modalities. The code is available at \href{https://github.com/idiap/mm-hsd}{https://github.com/idiap/mm-hsd}.

\noindent\textcolor{red_oh}{\textbf{Warning:} some of the elements of the paper contain hate speech examples, which could be disturbing to some readers.}

\end{abstract}


\maketitle

\section{Introduction}
Hate speech (HS) is "{\it a speech or address inciting hatred or intolerance, especially towards a particular social group on the basis of ethnicity, religious beliefs, sexuality, etc.}"~\cite{https://doi.org/10.1111/japp.12648}. 
The widespread use of social media and online fora~\cite{kemp2025state}, where people express their opinions on diverse subjects, has led to an increase in HS online~\cite{wu_detection_2020}. This proliferation of hate-related posts distorts political discourse, negatively affects public dialogue~\cite{AdvancesHS}, and can lead to the radicalization of individuals, increasing the risk of hate-related terrorism~\cite{hatereviewchallengessolutions}. Historically, hate speech detection (HSD) was performed manually~\cite{hatereviewchallengessolutions}, limiting its scalability and imposing a significant psychological burden on the moderators~\cite{wilson2020hate}.
With online content growing in volume and complexity, automated HSD reduces the need for human moderation by serving as an initial filter~\cite{rawat2024hate, aidmoderators}.

Since online HS has traditionally been associated with textual content~\cite{hee-etal-2024-recent}, text-based HSD has been extensively studied~\cite{caselli2020hatebert, Detoxify, mathew2021hatexplain}. However, social media content is increasingly multi-modal, and HS can appear not only in text but also in visuals and audio. To account for this diversity, recent work in multi-modal HSD incorporates multiple data sources, such as images~\cite{yang-etal-2019-exploring-deep, sandulescu2020detectinghatefulmemesusing, zhang2020hatefulmemesdetectioncomplementary} and user metadata~\cite{10.1145/3340531.3411934, hatereviewchallengessolutions}. While interest in multi-modal HSD is growing, it remains relatively underexplored compared to text-based approaches. Furthermore, most research on multi-modal HSD has focused on integrating images as an additional modality -- particularly in the context of memes and social media posts~\cite{suryawanshi-etal-2020-multimodal, HOSSAIN20226605, gomez2020exploring, mti5070034, caselli2020hatebert}. In contrast, work on multi-modal detection of HS in videos is relatively scarce, despite the rise of video-centric platforms like YouTube, Instagram, and TikTok, which facilitate its spread.

Video-based HSD is particularly challenging because hateful content may be embedded in multiple modalities, including video frames, on-screen text, and audio~\cite{das2023hatemm,chhabra2023literature}, often concealed within memes, music, or other non-traditional formats~\cite{jubany2016backgrounds}. Some prior work relies solely on audio transcripts~\cite{alcantara2020offensive, wu_detection_2020}, overlooking other modalities. Other studies integrate audio, video frames, and transcriptions~\cite{das2023hatemm, wang2024multihateclip,towardshatemmcomp}, but ignore the visual text within the frames, which could provide significant cues to improve detection accuracy. To our knowledge, there is only one study that incorporates on-screen text as a modality in HSD~\cite{xiong2024tcehatemmcomp}. 

Motivated by its strong performance in video classification~\cite{Praveen_2024_CVPR, gorti2022x}, in this paper, we explore Cross-Modal Attention (CMA)~\cite{chi2019two, madukwe2022token} as a flexible and context-aware fusion mechanism. Our insight is that, as CMA allows one modality to attend to another, it is especially useful for identifying necessary contextual cues when HS appears in a different modality. To this end, we propose to use CMA as a feature extractor for multi-modal integration at an early stage. 

Our main contributions are:
\begin{itemize} 
\item \textbf{Multi-modal HSD in videos}: We contribute to the limited literature on video HSD, developing a model that processes and fuses multiple features across modalities -- transcript, audio, video frames, and on-screen text. This multi-modal approach enables a more comprehensive representation of hateful content, especially when individual modalities alone are insufficient or ambiguous.

\item \textbf{CMA as early fusion for contextual integration:} We are the first to use CMA as an early fusion mechanism in HSD for videos, whose output is subsequently concatenated with modality-specific representations in a late fusion step before final prediction.

\item \textbf{Incorporating on-screen text as a standalone modality in CMA:} We are the first work to use on-screen text as a standalone channel in the context of CMA, showing that on-screen text attending to concatenated transcript, audio, and video frames features acts as a useful additional feature in the context of HSD.

\item \textbf{CMA-based interactions between the modalities: } We perform an analysis of the interactions between different modalities in the CMA setup, by testing different key-query combinations. We show that on-screen text performs best when used as a query attending to other modalities.

\end{itemize}

We release MM-HSD as an open-source benchmark for video-based HSD to support and advance ongoing research in this area\footnote{https://github.com/idiap/mm-hsd}.

\section{Related Work}

\noindent \textbf{HSD.} Early social media platforms primarily supported textual content, which led to the development of the first HSD methods in the textual domain (HSD-t). They relied on keyword-based approaches, using predefined dictionaries of offensive words~\cite{Hatebase, saleem2017web}. Although these methods could achieve high precision for explicit slurs, they suffered from low recall, since HS often depends on context~\cite{hatereviewchallengessolutions}. Also, due to the volume of online content, manual moderation is unscalable. To address these limitations, machine learning methods such as Naive Bayes classifiers~\cite{naivebayes} and Support Vector Machines (SVMs)~\cite{svmhsd} gained popularity~\cite{alrehili2019automatic}. Later, deep learning architectures like Convolutional Neural Networks (CNNs) and Long Short-Term Memory (LSTM) networks further improved performance~\cite{gandhi2024hate, roy2020framework}. Today, transformer-based models are the standard in HSD~\cite{roberta, mathew2021hatexplain, caselli2020hatebert, Detoxify, das2023hatemm}.
\begin{table}[t!]
\centering
\small
\caption{HSD in videos. KEY -- T:~text modality (speech transcript), A:~audio modality (waveform), V:~video modality (frames), O:~on-screen text (text in the frame), M:~video metadata, n/a: not applicable, MoE:~Mixture of Experts, MPL:~Multimodal Projection Layer in Llama 3.2-vision, CMA:~cross-modal attention, CONCAT:~concatenation fusion.}
\begin{tabular}{llllllc}
\hline
\multirow{2}{*}{\bf Reference} & \multicolumn{5}{c}{\bf Modality}                                                                         & \multirow{2}{*}{\bf Fusion} \\ 
                           & \multicolumn{1}{c}{T} & \multicolumn{1}{c}{A} & \multicolumn{1}{c}{V} & \multicolumn{1}{c}{O} & M &                         \\ \hline
                     \cite{alcantara2020offensive, wu_detection_2020} & \multicolumn{1}{l}{\blackcircle}  & \multicolumn{1}{l}{}  & \multicolumn{1}{l}{}  & \multicolumn{1}{l}{}  &   &            n/a             \\ \hline
                      \cite{Kandakatla2016IdentifyingOV} & \multicolumn{1}{l}{}  & \multicolumn{1}{l}{}  & \multicolumn{1}{l}{}  & \multicolumn{1}{l}{}  & \blackcircle  &              n/a           \\ \hline
                     
                     \cite{wang2024multihateclip, das2023hatemm, towardshatemmcomp} & \multicolumn{1}{l}{\blackcircle }  & \multicolumn{1}{l}{\blackcircle }  & \multicolumn{1}{l}{\blackcircle }  & \multicolumn{1}{l}{}  &   &    CONCAT                     \\ \hline
                     \cite{maity2024toxvidlmmultimodalframeworktoxicity, towardshatemmcomp} & \multicolumn{1}{l}{\blackcircle}  & \multicolumn{1}{l}{\blackcircle}  & \multicolumn{1}{l}{\blackcircle}  & \multicolumn{1}{l}{}  &   &        CMA                 \\ \hline

                    \cite{lang2025biting} & \multicolumn{1}{l}{\blackcircle}  & \multicolumn{1}{l}{\blackcircle}  & \multicolumn{1}{l}{\blackcircle}  & \multicolumn{1}{l}{}  & \blackcircle  &   MoE                      \\ \hline
                    \cite{crosshatemmcomp} & \multicolumn{1}{l}{\blackcircle}  & \multicolumn{1}{l}{}  & \multicolumn{1}{l}{\blackcircle}  & \multicolumn{1}{l}{}  &   &                  MPL       \\ \hline
                    \cite{xiong2024tcehatemmcomp} & \multicolumn{1}{l}{\blackcircle}  & \multicolumn{1}{l}{\blackcircle}  & \multicolumn{1}{l}{\blackcircle}  & \multicolumn{1}{l}{\blackcircle}  &   &  Bimodal CMA                       \\ \hline
                    MM-HSD (ours) & \multicolumn{1}{l}{\blackcircle}  & \multicolumn{1}{l}{\blackcircle}  & \multicolumn{1}{l}{\blackcircle}  & \multicolumn{1}{l}{\blackcircle}  &   &   CMA and CONCAT\\ \hline
\end{tabular}
\label{tab:related_work_modalities}
\end{table}

\noindent \textbf{Multi-modal Integration Techniques in HSD.} Multi-modal HSD integrates modalities such as images, audio, and metadata to improve detection performance~\cite{rawat2024hate}. Multiple modalities could be combined using different fusion strategies, such as, for example, concatenation~\cite{das2023hatemm, towardshatemmcomp, wang2024multihateclip}, and CMA~\cite{towardshatemmcomp, maity2024toxvidlmmultimodalframeworktoxicity, xiong2024tcehatemmcomp}. Although concatenation has an advantage in its implementation simplicity, it fails to capture inter-dependencies between the modalities~\cite{wu-etal-2021-multimodal}. CMA has gained relevance in HSD in videos (HSD-v)~\cite{towardshatemmcomp, xiong2024tcehatemmcomp}, and related domains such as toxicity, sexism and condescending language detection~\cite{maity2024toxvidlmmultimodalframeworktoxicity, arcos2024sexism, wang2024towards}, and in general video classification~\cite{chi2019two}, since it enables one modality to selectively focus on the most informative features of another~\cite{Praveen_2024_CVPR}. Its importance is highlighted in studies such as~\cite{madukwe2022token}, where the phrase “they shot another hamster” is benign, but “they shot another monkey” can be hateful if "monkey" is interpreted as a racial slur. This distinction can only be made when attending to both textual and visual modalities.

Multiple modalities can be combined with late or early fusion. Early fusion includes a combination of modalities directly before being fed into the decision model~\cite{duong2017multimodal, suryawanshi-etal-2020-multimodal}. Late fusion involves individual modalities to undergo high-level feature extraction and are then merged before passing through a final classification layer~\cite{vlad2020upb, ma2022hateful}. Early fusion focuses on fine-grained interactions between modalities, whereas late fusion focuses on coarse-grained interactions~\cite{pipoli_semantically_2025}.

\noindent \textbf{HSD in Videos.} HS in videos can be propagated via different channels, including spoken language, audio signals, visual content, and on-screen text. HSD-v methods differ based on which modalities are used, and how these modalities are combined (see Table~\ref{tab:related_work_modalities}). The first studies on HSD-v did not fully leverage multi-modal approaches but analyzed video transcripts~\cite{alcantara2020offensive, wu_detection_2020} or metadata such as YouTube titles, descriptions, and comments~\cite{Kandakatla2016IdentifyingOV}, which makes them essentially text-based approaches. 

Recent methods combine transcript (T), audio (A), and video~(V) using late fusion. Das et al.~\cite{das2023hatemm} showed that combining BERT, Vision Transformer (ViT), and MFCC improves Macro-F1 (M-F1) by 11.4\% over their best unimodal setup. Wang et al.~\cite{wang2024multihateclip} found that late fusion also outperformed across English and Bilibili datasets. Maity et al.~\cite{maity2024toxvidlmmultimodalframeworktoxicity} introduced CMA for HSD-v in Hindi-English code-mixed videos, using LLaMA-3, Whisper, and VideoMAE. CMA was only applied in bimodal (transcript-audio or transcript-video) settings, omitting audio-video interactions. Video and audio modalities are encoded separately and independently attend to text tokens, generating "soft multi-modal tokens" that are passed back into the language model. This setup is mid- to late-fusion and remains strictly text-centric, without exploring alternative query modalities. Optical Character Recognition (OCR) of on-screen text (O) is omitted entirely.

\begin{table*}[t!]
\centering
\small
\setlength{\tabcolsep}{2.5pt}
\caption{Performance comparison of models trained on HateMM \cite{das2023hatemm}. Bold indicates the best performance and underline the second-best for each metric. We report our results over 5 runs as mean (std). Key — T:~transcript, V:~video, A:~audio, O:~on-screen text, M:~macro average across both classes, H:~hate class, F1:~F1-score, ACC:~unbiased accuracy, P:~precision, R:~recall.}
\begin{tabular}{lllccccccccccc}
\hline
{\bf Model} & {\bf Architecture} &{\bf T}&{\bf V} &{\bf A} &{\bf O}
& {\bf ACC} & {\bf M-F1} & {\bf F1(H)} & {\bf P(H)} & {\bf R(H)} 
& {\bf P(M)} & {\bf R(M)} \\
\hline
\cite{das2023hatemm} & BERT, ViT, MFCC  & \blackcircle & \blackcircle & \blackcircle & 
& .798 & .790 & .749 & .742 & .758 
& -- & -- \\
\cite{towardshatemmcomp} & HXP, CLAP, CLIP  & \blackcircle & \blackcircle & \blackcircle & 
& \underline{.854} & \underline{.848} & -- & -- & -- 
& \underline{.840} & \underline{.800} \\
\cite{crosshatemmcomp} & LLaMA-3.2-11B  & \blackcircle & \blackcircle &  & 
& .820 & .820 & .800 & .800 & .790 
& -- & -- \\
\cite{xiong2024tcehatemmcomp} & BERT, ViT, wav2vec + OCR + CMA  & \blackcircle & \blackcircle & \blackcircle & \blackcircle
& .849 & .840 & \textbf{.876} & \textbf{.857} & \textbf{.896} 
& -- & -- \\
MM-HSD (ours) & Detoxify, ViT, wav2vec, OCR  + CMA & \blackcircle & \blackcircle & \blackcircle & \blackcircle 
& \textbf{.878 (.009)} & \textbf{.874 (.009)} & \underline{.853} (\underline{.009})& \underline{.849} (\underline{.017}) & \underline{.857} (\underline{.000})
& \textbf{.874 (.010)} & \textbf{.875 (.008)} \\
\hline
\end{tabular}
\label{tab:relatedresults_v2}
\end{table*}

Several models have been evaluated on the \textbf{HateMM} dataset~\cite{das2023hatemm}, with their performance summarized in Table~\ref{tab:relatedresults_v2}. HCC1~\cite{towardshatemmcomp} sets the state-of-the-art M-F1 (0.848) and accuracy (0.854), combining HateXplain (T), CLIP (V), and CLAP (A) via late fusion. Authors~\cite{towardshatemmcomp} also explore a variant with CMA (MO-Hate) with an attention chain, where the transcript attends to audio (T$\rightarrow$A), followed by text-audio attending to video ((T$\oplus$A)$\rightarrow$V). This chain underperforms in comparison to a simple concatenation, leading the authors to discard CMA. Moreover, it excludes on-screen text and only models bimodal interactions. The authors also report that Whisper significantly improves performance -- a component we adopt as well. TCE-DBF~\cite{xiong2024tcehatemmcomp} achieves the highest Micro-F1 (hate) score of 0.876, using CMA to combine text (transcript and on-screen), audio, and video — making it, alongside our work, the only model to incorporate on-screen text. The transcript and on-screen text are jointly processed using BERT without explicit modality signals, which may blur their semantic roles. In addition to this, modality interactions are limited: each modality is independently encoded (no early fusion), and text acts as the sole query in two distinct late-fusion cross-attention blocks (T$\rightarrow$A and T$\rightarrow$V). Also, alternative query-key combinations are not explored.

Vid+RM-FT~\cite{crosshatemmcomp} fine-tunes LLaMA-3.2-11B and LLaVA-Next-Video-7B on HateMM and a re-annotated version of the Hateful Memes dataset~\cite{kiela2020hateful}. However, it excludes on-screen text and relies on a general-purpose backbone not optimized for HSD-t, increasing the risk of overfitting due to the dataset's limited size. Lang et al.~\cite{lang2025biting} introduce a Mixture-of-Experts (MoE) model, where modality-specific experts are combined with weights predicted by a router conditioned on all modalities. However, the fusion mechanism may be too simplistic to fully capture cross-modal interactions. Additionally, as in earlier work, BERT is fine-tuned and OCR is omitted — both limitations that may impact performance. 

Prior CMA-based approaches use a single configuration without extensive comparison between possible setups. We are the first to evaluate CMA for HSD with early vs late fusion, query-key combinations, and modality integration/exclusion, which provides guidance for CMA design. Previous work \cite{towardshatemmcomp,maity2024toxvidlmmultimodalframeworktoxicity,xiong2024tcehatemmcomp} all rely on late fusion attention, and use text exclusively as query, omitting exploration of alternative modalities' interactions. Only TCE-DBF~\cite{xiong2024tcehatemmcomp} includes OCR by merging its output with the transcript . Koushik, Kanojia, and Treharne~\cite{towardshatemmcomp} attempted sequential CMA, but achieved poor performance, which made them discard CMA.

Finally, HS in videos datasets~\cite{das2023hatemm, wang2024multihateclip} include frame-level annotations, but all their associated models~\cite{towardshatemmcomp, crosshatemmcomp, xiong2024tcehatemmcomp} are video-level. A (partial) exception is~\cite{lang2025biting} that segments videos for relevance scoring, but outputs a video-level prediction. Related unimodal work includes audio-level localisation via TTS~\cite{kibriya2024towardsxaitext} and word-level localisation~\cite{an2024investigation}. No multi-modal HS models explore true frame-level localisation~\cite{lang2025biting, maity2024toxvidlmmultimodalframeworktoxicity}.

\section{Dataset}

Contrasting with the abundance of textual HS datasets~\cite{caselli2020hatebert, mathew2021hatexplain, davidson2019racial, de2018hate}, video-based datasets remain underdeveloped~\cite{wang2024multihateclip, lippe2020multimodal}. Datasets that claim to address video-based hate-speech detection are often unimodal, containing only text comments or video transcriptions~\cite{gupta2023hateful, alcantara2020offensive, debele2022multimodal}. Many of these datasets are not publicly available~\cite{rana2022emotion, wu_detection_2020}\footnote{We did not receive a reply from the authors to get access to the datasets.}, or merely provide links to video platforms~\cite{wang2024multihateclip}.
Other datasets focus on related yet distinct concepts such as cyberbullying~\cite{ayetiran2024review}, sexism~\cite{arcos2024sexism}, or condescending language~\cite{wang2024towards}, and are therefore not wholly applicable to HS detection. Additionally, the applicability of some datasets is affected by the language in which HS is expressed: some datasets are in Hinglish (a mixture of English and Hindi)~\cite{maity2024toxvidlmmultimodalframeworktoxicity}, or in Bengali~\cite{hossain_junaid_bangla_2021}, and therefore might propagate HS differently to datasets in English. 

\begin{figure}[t]
    \centering
    \begin{minipage}[t]{0.48\linewidth}
        \centering
        \includegraphics[width=\linewidth]{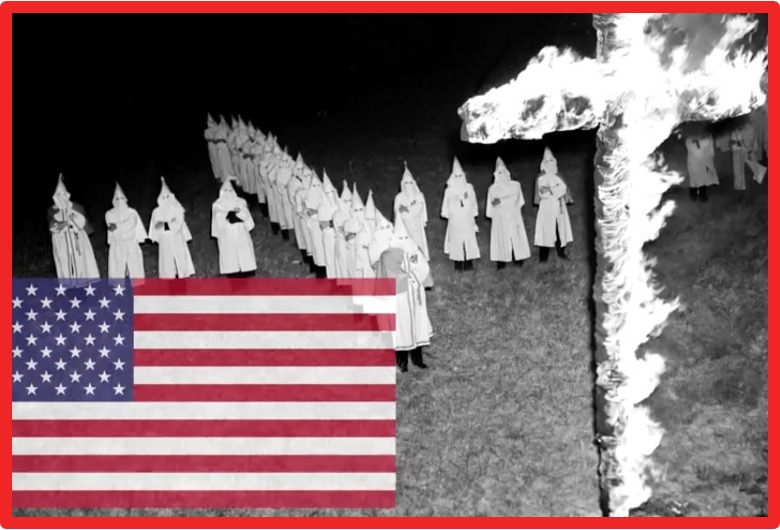}
        \scriptsize
        \parbox[t]{\linewidth}{(\textbf{a}) Transcript: \textit{They're marching for equality, they'll never be as good as me. We won't let them integrate, we must always segregate.}}
    \end{minipage}
    \hfill
    \begin{minipage}[t]{0.48\linewidth}
        \centering
        \includegraphics[width=\linewidth]{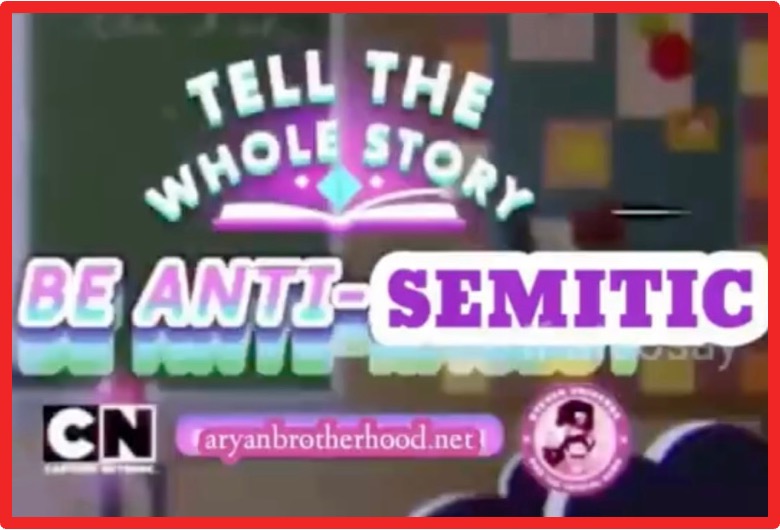}
        \scriptsize
        \parbox[t]{\linewidth}{(\textbf{b}) Transcript: \textit{Does six million sound like a real world estimate? [...] Do you honestly believe any of these Jewish lies?}}
    \end{minipage}


    \begin{minipage}[t]{0.48\linewidth}
        \centering
        \includegraphics[width=\linewidth]{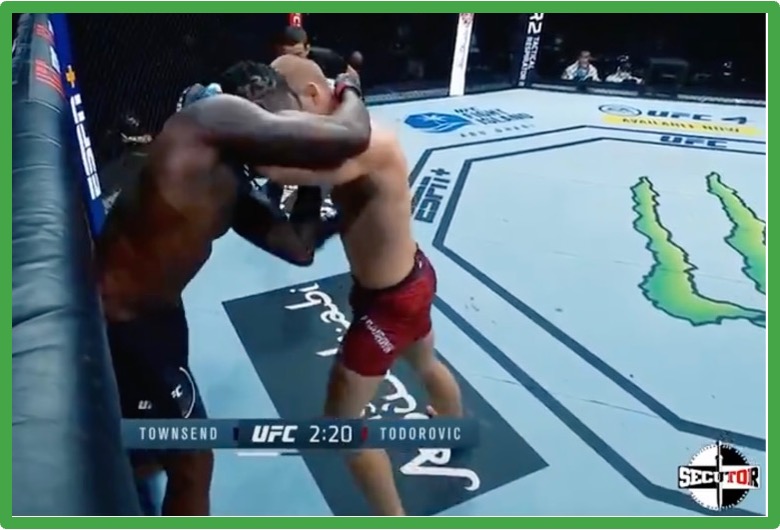}
        \scriptsize
        \parbox[t]{\linewidth}{(\textbf{c}) Transcript: \textit{None}}
    \end{minipage}
    \hfill
    \begin{minipage}[t]{0.48\linewidth}
        \centering
        \includegraphics[width=\linewidth]{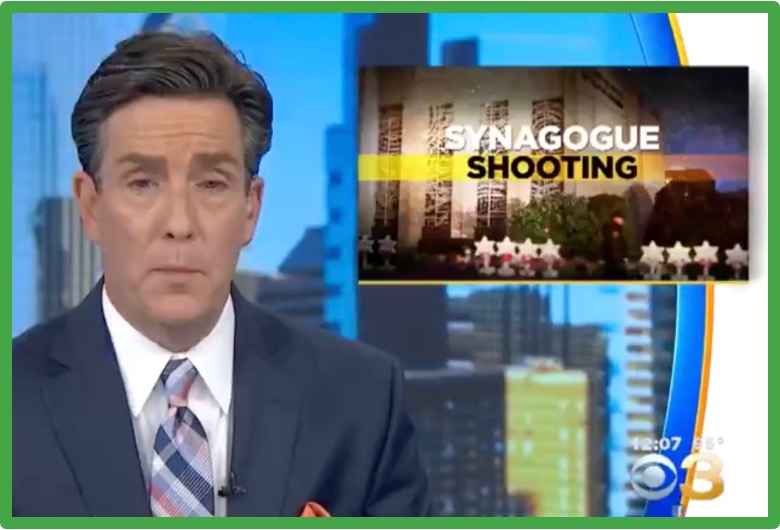}
        \scriptsize
        \parbox[t]{\linewidth}{(\textbf{d}) Transcript: \textit{The show up for Shabbat event is aimed at combating anti-semitism, hate and bigotry.}}
    \end{minipage}

    \caption{Sample frames extracted from videos classified as \textcolor{red_oh}{HATE} ((\textbf{a}), (\textbf{b})) and from videos classified as \textcolor{green_oh}{NO HATE} ((\textbf{c}),~(\textbf{d})). Frames adapted from HateMM dataset~\cite{hatemm_dataset}, CC BY 4.0.}
    \label{fig:all_images}
\end{figure}

In this study, we use the HateMM dataset, a publicly available dataset comprising 1083 labeled videos from the BitChute platform~\cite{das2023hatemm}, totalling 43 hours. This dataset contains 431 videos (39.8\%) labeled as hate and 652 (60.2\%) as non-hate. Video lengths vary significantly, ranging from a few seconds to over an hour, with an average duration of 2.40 minutes--2.56 minutes for hate videos and 2.28 minutes for non-hate videos. The dataset stands out for its diversity in how HS is represented, including videos where HS is conveyed through spoken words, displayed in text on video frames, or implied through actions depicted in the footage. Such a range is essential to underscore the potential of multi-modal models. In Figure~\ref{fig:all_images}, we present different examples of hate and non-hate videos from the dataset. Example (a) displays a Ku Klux Klan (Christian extremist, white supremacist, far-right hate group) gathering with a voiceover stating "We must segregate". While the image alone might have documentary purposes, the combination of visual and audio elements is unambiguously hateful. In contrast, example (b) features on-screen text explicitly calling to antisemitism, constituting HS through the text in the visual modality. This demonstrates the importance of including OCR of video frames into the model, enabling the detection of HS in on-screen text that might be missed by image-only analysis. On the other hand, examples (c) and (d) show images that could be interpreted as HS if there was an accompanying hateful text or voiceover. However, (c) is a clip from a mixed martial arts fight and (d) features news reports; both serve a purely informative purpose.

Additionally, the dataset comprises a variety of video types, which enhances its applicability and generalizability to real-world scenarios. Some videos consist of static images overlaid with text, accompanied by voiceovers or background music, while others include dynamic footage of individuals in public or private spaces. 
\section{Methodology}
\label{sec:methodology}

\begin{figure*}[t]
    \centering
\includegraphics[width=0.9\linewidth]{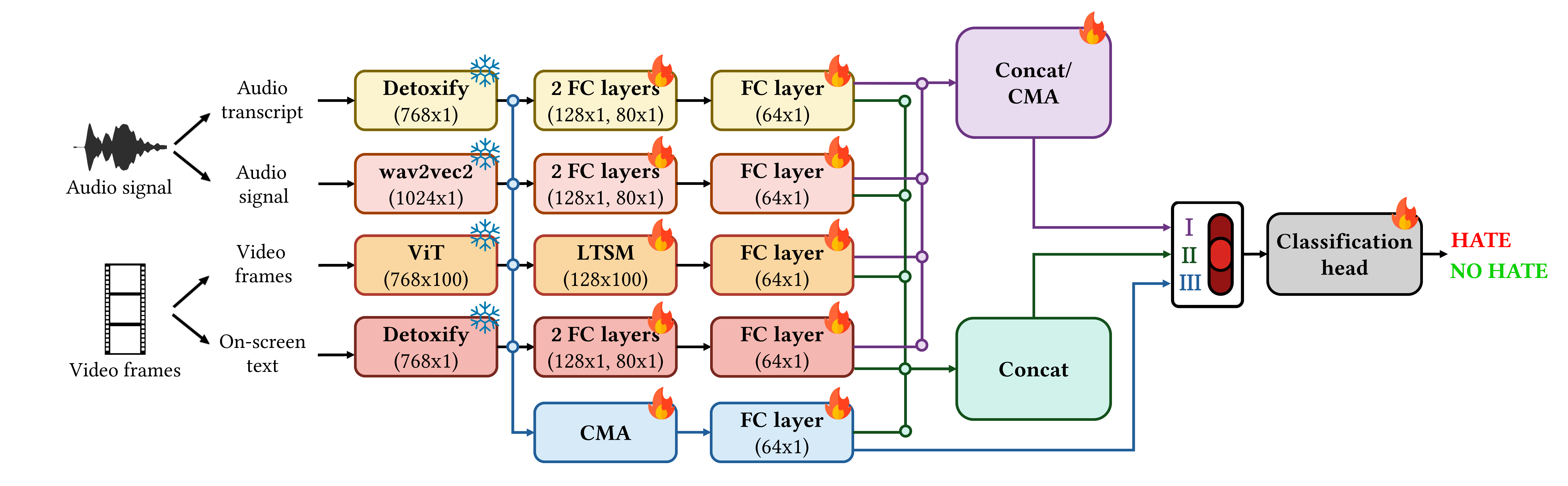} 
    \caption{Our tetra-modal HSD architecture. The four 
    modalities are audio transcripts, audio signal, video frames, and on-screen text. The embeddings for each modality are obtained by pre-trained models and then used in three different experimental setups, controlled by a switch module with three configurations (I, II, III). \fcolorbox{white}{lavender}{I) Late Fusion Experiments}: Each modality is first independently encoded (e.g.~by LSTM or Fully Connected (FC) layers), and their outputs are fused using either concatenation or CMA.\fcolorbox{white}{mint}{II) Late Fusion with CMA as Additional Modality}: CMA is applied to raw modality embeddings to generate an additional feature, which is then combined (via concatenation) with the encoded features. \fcolorbox{white}{lightblue}{III) Early Fusion with CMA as Unique Feature Extractor}: Raw modality embeddings are fused directly using CMA, without any modality-specific processing. 
     The final classification head outputs a binary label: \textcolor{green_oh}{NO HATE} or \textcolor{red_oh}{HATE}.}
    \label{fig:diagram_unified}
\end{figure*}

This section presents our approach for developing the MM-HSD multi-modal model. We begin by discussing the CMA mechanism~\cite{chi2019two} and how we propose to integrate CMA as an extra modality. Assuming the availability of pre-extracted embeddings for each modality (video, audio, transcript, and on-screen text), we then describe how CMA is integrated into the overall multi-modal architecture. This is followed by a comprehensive explanation of the embedding extraction process. 
Since the availability of video datasets for HSD is currently very limited, it is inefficient to fine-tune unimodal feature extractors (such as BERT) based on these small multi-modal datasets, as done in previous works~\cite{wang2024multihateclip, wang2024towards}. For this reason, we use already fine-tuned HS recognition models — such as Detoxify~\cite{Detoxify} — as text feature extractors, since they have been trained on much larger datasets for HSD-t. A detailed diagram of the complete pipeline is presented in Figure~\ref{fig:diagram_unified}.

\subsection{Models}
\label{sec:model_arch}
\noindent \textbf{CMA.} CMA integrates information from different modalities by allowing the model to focus on specific aspects of one modality that are most relevant to another~\cite{chi2019two}. We explore the contribution of CMA by primarily evaluating it as an additional processing module output of which is fused with the modality-specific model outputs (see Figure~\ref{fig:diagram_unified}, II) -- this constitutes our MM-HSD model. We additionally evaluate its performance in two more distinct roles, which are illustrated in Figure~\ref{fig:diagram_unified} with color-coded paths: I) a late fusion layer that integrates the outputs of modality-specific models for video, audio, transcript and on-screen text, which we call CMA-LF, and III) a standalone feature extractor that directly processes the raw embeddings from all modalities without using separate models for each modality, CMA-S. The goal is to further incorporate relationships between different modalities~\cite{liu2021cma}. 

CMA consists of an attention mechanism with Query (Q), Key~(K), and Value (V). It is expressed as 
\[
\text{CMA}(K, Q, V) = \text{softmax} \left( \frac{QK^T}{\sqrt{d_k}} \right) V,
\]
where \( d_k \) is the dimensionality of the key vectors. In this work, Q represents the modalities that extract relevant information from other modalities. K are the modalities to which the query attends to identify the most relevant information. V holds the actual data, from which the most relevant parts are extracted based on the query-key interaction. The choice of Q, K, and V values is studied, and thus different combinations are tuned to find the optimal combination. The K and V modalities are identical, while the Q is assigned to the remaining modality that is not used for K and V. 
For instance, in a transcript-video-audio setup, some of the scenarios in which the model is tested include: 
\begin{itemize}
    \item \textbf{K}=[audio, transcript], \textbf{Q}=[video], \textbf{V}=[audio, transcript]
    \item \textbf{K}=[video, transcript], \textbf{Q}=[audio], \textbf{V}=[video, transcript]
    \item[] ...
\end{itemize}

Note that in the case of multiple modalities serving as the key, they are concatenated along the sequence dimension. This means that the embeddings from different modalities are stacked sequentially, allowing the attention mechanism to process them as a unified extended sequence rather than treating each modality separately~\cite{towardshatemmcomp}.

\noindent \textbf{Model Ensembling with CMA as an extra modality (MM-HSD).} The pipeline for MM-HSD is described in Figure~\ref{fig:diagram_unified}, II. CMA is applied directly to the raw modality embeddings, and its output is concatenated with the outputs of the individual modality encoders before a final classification head. We use concatenated features of T, V, and A modalities as keys, keeping O as a query, a combination that was experimentally found to yield the best performance (see Appendix~A, Table~7). An analysis of the selection process is presented in Section~\ref{sec:last}. We report the performance of MM-HSD in Table~\ref{tab:relatedresults_v2}.

\noindent \textbf{CMA as a standalone feature extractor (CMA-S).}
We apply CMA directly to the raw modality embeddings. We evaluate the predictive power of CMA in the early fusion settings (see Figure~\ref{fig:diagram_unified},~III). The output of CMA is then passed through a feedforward layer for final classification, with no additional unimodal transformations. We use concatenated features of T, V, and A modalities as keys, and O as a query. 

\noindent
\textbf{CMA as a late fusion strategy (CMA-LF).} CMA is used as a late fusion strategy, on the outputs of the individual modality encoders before the final classification head (see Figure~\ref{fig:diagram_unified}, I). 

\subsection{Embeddings extraction and encoding}
We now describe the pre-processing and feature extraction steps for each individual modality, followed by the subsequent encoding stage.

\noindent \textbf{Video.}
Video files are sampled at one frame-per-second, with a maximum of 100 frames extracted per video, following~\cite{towardshatemmcomp, das2023hatemm}. Shorter videos are padded with blank frames to reach 100 frames. The embeddings are generated using a pretrained ViT ~\cite{dosovitskiy2021image}. We apply the ViT to each frame independently and concatenate their embeddings. ViT has been chosen for visual feature extraction due to its ability to model long-range dependencies and thus capture global context through self-attention mechanisms -- unlike convolutional networks, which rely on local receptive fields. In addition, we aim to separate the concerns of image context contribution and on-screen text cues. Newer models like CLIP~\cite{clip} have shown to have higher sensitivity towards on-screen text than to the image components, such as texture~\cite{noever2021reading}, while ViT should be able to focus on the context within the image rather than on the on-screen text.

\noindent \textbf{Audio and Transcriptions.} From each video, we extract its audio, which we then transcribe. This way we separate the content of the human speech, which is contained in the transcript and could be processed by a pre-trained hate-speech encoder, from the emotional cues contained in the speech, such as aggressiveness, which are contained in audio~\cite{das2023hatemm}. To obtain acoustic features, we use a pre-trained wav2vec2-large-xlsr-53 model~\cite{baevski2020wav2vec20frameworkselfsupervised, grosman2021xlsr53-large-english}, a self-supervised speech recognition model composed of a convolutional feature encoder and a transformer network fine-tuned for the English language. Finally, we transcribe the audio using OpenAI's Whisper model~\cite{radford2023robust}. This text is then encoded into low-dimensional vectors using Detoxify~\cite{Detoxify}, a model based on RoBERTa~\cite{roberta} and specifically trained to detect various forms of toxic and hateful language. 
A complementary ablation on removing stopwords on audio transcripts is included in Appendix~B. Stopwords removal led to a performance drop and was not used in the main experiments.

\noindent \textbf{On-screen text.} To extract the printed text, captions, or other textual elements within the video frames, we sample one frame per second. The text from each frame is extracted using PaddleOCR~\cite{paddleocr}, since it can handle different text orientations and noisy frames with minimal postprocessing. Once extracted, text coherence is enhanced by applying the following postprocessing steps. First, the text is cleaned by removing any unwanted characters, retaining only alphanumeric symbols, common punctuation, and apostrophes for contractions. Next, de-duplication is performed to reduce redundancy introduced by frames extracted at short intervals. Duplicate or highly similar text segments are filtered out by comparing their similarity scores and discarding those over 90\% resemblance to an already retained entry. Finally, overlapping text fragments are merged by identifying the longest matching sequences and combining them into a cohesive output, ensuring a more accurate and readable final text. Similarly to the audio transcriptions, we embed the OCR output with a Detoxify model.

We tested the effect of replacing MM-HSD’s original feature extractors one modality at a time — using BERT instead of Detoxify for transcript and on-screen text, MFCC instead of wav2vec for audio, and InceptionV3 instead of ViT for video. These alternative extractors underperformed compared to the chosen ones. For instance, MM-HSD outperforms all alternatives by 2.4–4.7\% in M-F1. This supports the chosen extractors for each modality.

\noindent \textbf{Encoding of individual modalities.} This component is used in setups I and II. The video modality is represented by 768-dimensional features extracted from each frame, which are fed into an LSTM network, followed by an FC layer for hate video classification. LSTM networks excel at detecting patterns and dependencies over time, such as actions within the video stream, making it a suitable choice for understanding the dynamics between frames. For text, we input the 768-dimensional embeddings from both the audio and OCR transcripts into separate neural networks, each consisting of three FC layers. For the audio signal, the 1024-dimensional wav2vec2 features serve as input for training a three-FC layer neural network. 

\section{Validation}

\subsection{Experimental setup}
Given a limited amount of data, we adopt a 5-fold cross-validation strategy on 85\% of the data, reserving the remaining 15\% for testing. Each fold includes 698 training and 175 validation samples, with a batch size of 8. All runs are executed on an NVIDIA RTX 3090. We employ elastic net regularization, combining L1 and L2 penalties to enhance sparsity and stability. A ReduceLROnPlateau scheduler reduces the learning rate by a factor of 0.1 when validation loss does not decrease for 6 consecutive epochs. Early stopping is triggered if no improvement is observed over a patience period. Dropout is applied after each FC layer to further regularize the model. To account for class imbalance, we use a weighted cross-entropy loss. Hyperparameter tuning is performed over a fixed grid, with the learning rate \(\eta \in \{10^{-3}, 10^{-4}, 10^{-5}\}\), L1 penalty \(\lambda_1 \in \{10^{-3}, 10^{-4}, 10^{-5}\}\), L2 penalty \(\lambda_2 \in \{10^{-4}, 10^{-5}, 10^{-6}\}\), dropout rate \(d \in \{0.3, 0.4, 0.5\}\), and early stopping patience \(p \in \{5, 10\}\). 

\noindent \textbf{Metrics.} Model performance is assessed using both overall and class-specific metrics. Unbiased Accuracy (ACC) -- the mean of class-wise recall in binary classification -- ensures performance is not skewed by class imbalance. Hate F1 (F1(H)) captures performance specifically on the Hate class, while M-F1 averages F1 scores across both Hate and No Hate classes. We additionally report Hate Precision (P(H)) and Hate Recall (R(H)). The best-performing models are chosen according to the training and validation loss trends and the corresponding validation F1, recall, and precision macro-averaged and hate-specific scores. In Table~\ref{tab:relatedresults_v2}, we summarize the comparative performance of state-of-the-art models on HateMM~\cite{das2023hatemm} dataset. Note that while TCE-DBF~\cite{xiong2024tcehatemmcomp} reports higher hate metrics, our model outperforms in macro metrics. Additionally, results of~\cite{xiong2024tcehatemmcomp} are from a single run with unknown variability. 
\begin{table}[t]
\centering
\small
\setlength{\tabcolsep}{2.5pt}
\caption{Results for unimodal experiments models, CMA as a standalone feature extractor (CMA-S), CMA as an extra modality (MM-HSD), CMA as late fusion (CMA-LF) and modality-specific models fused with concatenation without CMA (w/o CMA). 
We report our results over 5 runs as mean (std). Key — T:~transcript, V:~video, A:~audio, O:~on-screen text, M:~macro average across both classes, H:~hate class, F1:~F1-score, ACC:~unbiased accuracy, P:~precision, R:~recall.}
\begin{tabular}{lccccc}
\hline
{\bf Model} & {\bf ACC} & {\bf M-F1} & {\bf F1(H)} & {\bf P(H)} & {\bf R(H)} \\
\hline
T & .820 (.012) & .816 (.012) & .790 (.012) & .765 (.019) & .816 (.009) \\
O & .636 (.014) & .594 (.011) & .464 (.012) & .596 (.032) & .381 (.016) \\
A & .784 (.019) & .778 (.018) & .742 (.018) & .739 (.039) & .746 (.030) \\
V & .761 (.027) & .751 (.024) & .702 (.020) & .730 (.055) & .679 (.017) \\
CMA-S$^{\dagger}$ & .850 (.006) & .846 (.006) & .820 (.006) & .818 (.016) & .821 (.008) \\
MM-HSD$^{\dagger}$ & \textbf{.878 (.009)} & \textbf{.874 (.009)} & \textbf{.853 (.009)} & \textbf{.849 (.017)} & \textbf{.857 (.000)} \\
w/o CMA & .846 (.013) & .842 (.014) & .817 (.019) & .805 (.028) & .832 (.052) \\
CMA-LF$^{\dagger}$ & .842 (.024) & .837 (.024) & .810 (.028) & .812 (.057) & .813 (.057) \\
\hline
\end{tabular}
\vspace{0.3em}
\raggedright\scriptsize
$^{\dagger}$ Query: On-screen text. Key: concatenation of transcripts, video, and audio.
\label{tab:unimodal}
\end{table}

\subsection{Baseline models comparison}
\label{sec:baseline}

We evaluate the predictive strength of the individual modalities of our model.  
In the first rows of Table~\ref{tab:unimodal} we present the unimodal baselines (T, V, A, O), and the CMA as a standalone feature extractor baseline (CMA-S), which processes the four modalities in the early fusion setting, as described in Section~\ref{sec:model_arch}. 
Consistent with the literature~\cite{das2023hatemm, maity2024toxvidlmmultimodalframeworktoxicity}, we observe that the T modality is the most effective unimodal model, achieving an M-F1 score of 0.816. In contrast, the O modality yields the lowest performance, with an M-F1 score of 0.594. This disparity is expected, as not all videos in the dataset contain on-screen text, making the O modality ineffective in some instances. CMA-S uses the O modality as the query and the T, A, and V modalities as keys. The performance of CMA-S being higher (M-F1 of 0.846) than unimodal baselines is to be expected, as it already constitutes a fully multi-modal model that combines information from all modalities. As such, it already reflects the benefits of multi-modal HSD and the advantages of using CMA.

We also compare the CMA mechanism used independently as a feature extractor, CMA-S, with CMA incorporated as an additional modality within the full multi-modal framework, MM-HSD. As shown in Table~\ref {tab:unimodal}, combining CMA and the individual models results in a boost in performance, increasing M-F1 score from 0.846 to 0.874. The reason behind this is that leveraging modality-specific encoders allows the model to use both cross-modal interaction and the specialized representations captured by each unimodal encoder. Note that we maintain the query-key pair aforementioned, with O as query and TVA as key for both CMA-S and MM-HSD. For completeness, we evaluate CMA as a late fusion strategy for the modality-specific models, CMA-LF, which underperforms (M-F1 score of 0.837) compared to MM-HSD. We use MM-HSD for the rest of the experiments.

\subsection{Decreasing the number of modalities}

\begin{table}[t]
\centering
\small
\setlength{\tabcolsep}{2pt}
\caption{Results for CMA as an extra modality (MM-HSD). We report our results over 5 runs as mean (std). Key~--~Mod:~modalities, K:~key, Q:~query, T:~transcript, O:~on-screen text, A:~audio, V:~video, M:~macro average across both classes, H:~hate class, F1:~F1-score, ACC:~unbiased accuracy, P:~precision, R:~recall.}
\begin{tabular}{l@{\hskip 1.5pt}l@{\hskip 1.5pt}lccccc}
\hline
{\bf Mod.} & {\bf K} & {\bf Q} & {\bf ACC} & {\bf M-F1} & {\bf F1(H)} & {\bf P(H)} & {\bf R(H)} \\
\hline
TO   & T   & O   & .830 (.006) & .825 (.005) & .796 (.006) & .793 (.014) & .800 (.014) \\
TA   & T   & A   & .828 (.025) & .823 (.024) & .796 (.023) & .786 (.046) & .806 (.007) \\
TV   & T   & V   & .841 (.006) & .837 (.006) & .811 (.006) & .799 (.009) & .822 (.007) \\
OA   & A   & O   & .805 (.028) & .801 (.027) & .774 (.024) & .749 (.048) & .803 (.014) \\
OV   & V   & O   & .775 (.009) & .768 (.010) & .730 (.014) & .726 (.005) & .733 (.026) \\
AV   & A   & V   & .808 (.026) & .799 (0.030) & .759 (.041) & .788 (.023) & .733 (.065) \\
TOA  & TO  & A   & .834 (.011) & .830 (.011) & .805 (.014) & .787 (.023) & .825 (.037) \\
TOV  & TV  & O   & .838 (.019) & .834 (.019) & .807 (.022) & .800 (.035) & .816 (.040) \\
TVA  & TV  & A   & .849 (.007) & .845 (.007) & .819 (.009) & .811 (.014) & .829 (.021) \\
OAV  & OA  & V   & .821 (.023) & .815 (.026) & .781 (.035) & .789 (.013) & .775 (.059) \\
TOAV & TAV & O   & \textbf{.878 (.009)} & \textbf{.874 (.009)} & \textbf{.853 (.009)} & \textbf{.849 (.017)} & \textbf{.857 (.000)} \\
\hline
\end{tabular}
\label{tab:decreasing_modalities}
\end{table}

\begin{figure}
    \centering
    \includegraphics[width=0.77\linewidth]{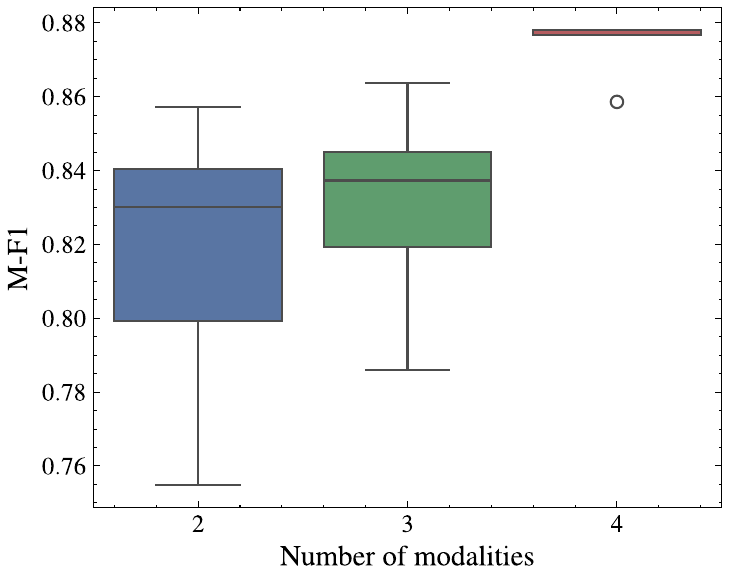}
    \caption{Impact of modality combinations on the M-F1 score. Each box represents the distribution of mean M-F1 scores (mean computed over five runs with different random seeds) for each specific combination of modalities, grouped by the number of modalities used — from bimodal to tetra-modal setups. The combinations include all possible subsets of the four modalities (T, A, V, and O) with key-query pairs selected on training and validation metrics on CMA-S. The trend shows that M-F1 generally increases as more modalities are integrated.}    \label{fig:Modalities_M-F1}
    \vspace{-10pt}
    \end{figure}

In Table~\ref{tab:decreasing_modalities}, we show how model performance changes when individual modalities (T, O, A, V) are removed from the full tetra-modal setup of MM-HSD. The key-query pairs used are chosen according to the training and validation metrics obtained when using CMA-S. 
When dropping any single modality from the tetra-modal configuration (\mbox{M-F1} = 0.874), performance declines to between 0.815 and 0.845, already showing that all modalities contribute unique information for HSD. The largest drop occurs when the T modality is removed, and the lowest when the O modality is removed. Likewise, in the transition from a trimodal system (best \mbox{M-F1}~=~0.845) down to any bimodal setup, scores drop further, between 0.768 and 0.837. The same modalities that drive performance in the tetra-modal setup also contribute the most when fewer modalities are used. Reducing the number of input modalities increases the range of the mean M-F1 scores, which is illustrated in Figure \ref{fig:Modalities_M-F1}. We further observe that the specific combination of modalities used as keys becomes more influential -- leading to substantial variation within 2-, 3-, and 4- modality settings. 

To further isolate the contribution of each modality within the attention mechanism, we perform an additional experiment on MM-HSD in which modalities are selectively excluded from the CMA block only, with the O modality fixed as the query. The resulting attention output is then concatenated with the individual modality encoder outputs before classification, following the original architecture. We observe consistent drops in M-F1 and F1(H) when any modality is excluded (see Appendix~C), further confirming the need to include all modalities.

\subsection{Removing CMA features as a modality}

To assess the contribution of the CMA mechanism to overall model performance, we evaluate the configuration with only the modality-specific encoders, and excluding CMA as a feature extractor. This enables a clearer understanding of the added value provided by CMA beyond the capabilities of the individual modalities in isolation. For this comparison, we leverage concatenation as the fusion technique. We observe that incorporating CMA as an extra modality results in a performance boost: M-F1 score of 0.878 (MM-HSD) against 0.846 (w/o CMA), as shown in Table \ref{tab:unimodal}. Moreover, the incorporation of CMA leads to a significant reduction in standard deviation, indicating improved consistency and robustness across evaluation runs.

\subsection{Performance trends by CMA configuration}
\label{sec:last}

We evaluate the relationship between different modalities using CMA as a fusion strategy in two experimental setups: CMA-S and CMA-LF. In these configurations, the difference lies in the input to CMA-S being raw embeddings, and the input to CMA-LF being encoded embeddings. 
In both configurations, we explore all possible query-key combinations where one modality (the query) attends to a set of other modalities (the keys), which are concatenated together. Conceptually, when a modality \( A \) attends to \( \text{concat}(B_1, ..., B_n) \), this reflects the amount of useful information modality \( A \) can extract from the modalities it attends to. The full list of key-query pairs used in these experiments is provided in Appendix~A, Tables 6 and~7. Note that these experiments were conducted using a single-seed setup.

\begin{figure}[t]
    \centering
\includegraphics[width=0.9\linewidth]{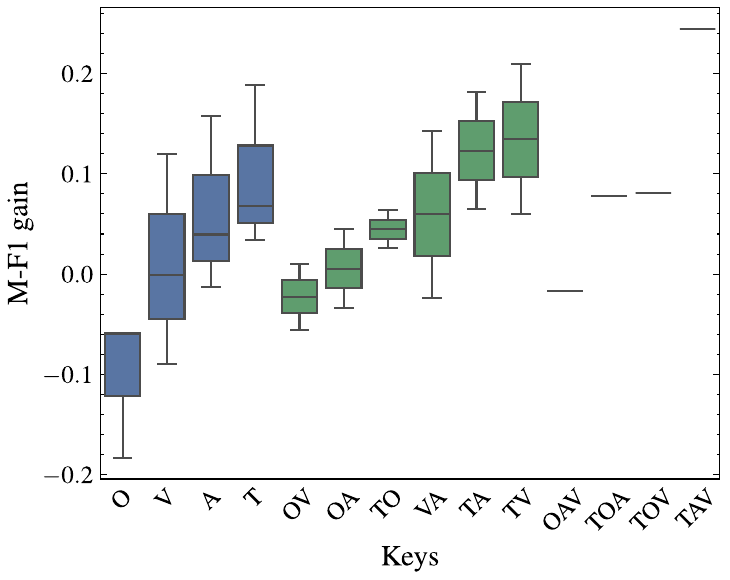}
    \caption{The dependence between keys used in CMA fusion and the average M-F1 performance gain in comparison to a corresponding unimodal baseline (model trained on the query modality). For each key, the distribution includes all possible query options that do not share any modalities with that key. The distribution is based on the average M-F1 gain for CMA-S and CMA-LF. Key -- T: transcript, V: video, A: audio, O: on-screen text, M-F1: Macro-F1 score.}
    \label{fig:avg_gain_keys}
    \vspace{-10pt}
\end{figure}

To evaluate the impact of these interactions, we compare the M-F1 scores of CMA-S and CMA-LF against those of the unimodal baselines, where the modality used in the unimodal model matches the query used in CMA. The gain in performance is then calculated as the difference in M-F1 between the CMA-based model and its unimodal counterpart.
First, we validate the stability of the results by examining the correlation between performance gains for both CMA-S and CMA-LF. We observe a very strong positive correlation of 91\%. This suggests that the patterns observed hold for both late and early fusion, and therefore can be generalized. To increase robustness, we take the average between the gain in performance for late and early setups, and conduct our analysis based on these averaged values (see Figure \ref{fig:avg_gain_keys}). 
We note that in general, whenever O is used as a part of a key, the performance drops, meaning that on-screen text does not allow enrichment of other modalities when used as a key. Using a transcript as a part of the key, allows the queries to gain a boost in performance, meaning that stronger modalities can often enrich the weaker modalities when used as a key. We note that the greatest gain happens when O is attending to T, A, and V. 

\subsection{Computational cost analysis}
We summarize the computational cost of our models in Table~\ref{tab:efficiency}. Training time per epoch is measured on 698 training and 175 validation samples, and the results are averaged over 5 folds. There is variation in training time across models introduced by early stopping. We observe that models with limited training capacity (e.g.~unimodal O baseline) complete training sooner. However, CMA runtime scales quadratically with input size, resulting in much longer training times for models such as CMA-S and \mbox{MM-HSD}. Inference time is calculated over 155 test samples, and remains consistently low across all models, even though using CMA greatly increases parameter count in models such as CMA-S and MM-HSD.

Note that feature extraction using Detoxify, ViT, and wav2vec2 is performed offline, and not counted towards the total runtime of the different models. These feature extractors are large (Detoxify contains 109M parameters,  ViT -- 86M, and wav2vec2 -- 315M) and not well-suited for on-device deployment. However, the final classifiers are relatively lightweight, with a maximum size of 4.6M parameters for MM-HSD, making on-device use possible if the extracted features are available.

\section{Conclusion}
\label{sec:conclusion}

We proposed MM-HSD, a model for multi-modal HSD in videos, and analyzed the contribution of audio, video, speech transcripts, and on-screen text modalities. We showed that incorporating CMA as an additional modality alongside modality-specific encoders leads to improved performance compared to both unimodal models and late-fusion strategies. Our results suggest that decoupling visual feature extraction from on-screen text recognition -- by using a general-purpose image encoder (e.g. ViT) in combination with a separate OCR module -- allows for a more nuanced HSD.
By analyzing query-key pairs in the context of CMA, we showed that on-screen text attending to transcripts, audio, and video yields the best performance. 

As future work, one could evaluate whether OCR-to-speech conversion improves HS classification within the CMA framework~\cite{AgarwalA, AgarwalB}.
Furthermore, it would be interesting to explore frame-level localization using temporal CMA~\cite{mercea2022temporal}, which may improve model explainability by showing which frames contributed to the final classification. Lastly, MM-HSD has been trained to optimize hate classification on HateMM solely. Future research should focus on generalization and validation across additional datasets.

\begin{table}[t]
\centering
\small
\setlength{\tabcolsep}{4pt}
\caption{Efficiency metrics for unimodal and multimodal models. Key -- CMA used as standalone (CMA-S), additional modality (MM-HSD), late fusion (CMA-LF), and removed (w/o~CMA), TTE: Train Time per Epoch, TTT: Total Train Time, TT: Test Time, Par: Parameters.}
\begin{tabular}{lccccc}
\hline
\multirow{1}{*}{\bf Model} & \makecell{\textbf{TTE} \\ \textbf{(s)}} & \makecell{\textbf{TTT}\\ \textbf{(s)}} & \makecell{\textbf{TT} \\ \textbf{(s)}} & \makecell{\textbf{\# Par} \\ \textbf{(M)}} & \makecell{\textbf{Size} \\ \textbf{(MB)}} \\
\hline
A        & 0.540 & 73.162 & 0.046 & 0.147 & 0.562 \\
T        & 0.441 & 65.818 & 0.058 & 0.123 & 0.470 \\
O        & 0.426 & 18.075 & 0.050 & 0.123 & 0.470 \\
V        & 0.462 & 31.427 & 0.041 & 1.279 & 4.880 \\
CMA-S    & 1.124 & 155.917 & 0.068 & 2.953 & 11.266 \\
MM-HSD   & 1.465 & 293.022 & 0.060 & 4.626 & 17.648 \\
w/o CMA  & 0.975 & 70.013 & 0.065 & 1.673 & 6.381 \\
CMA-LF   & 1.271 & 81.223 & 0.089 & 1.722 & 6.570 \\
\hline
\end{tabular}
\label{tab:efficiency}
\end{table}

\textit{}
\begin{acks}
P.R.-R. received the support of a fellowship from "la
Caixa" Foundation (ID 100010434), with fellowship code 
LCF/BQ/EU23/12010085. C.C.-C. received support of a fellowship from "Mutua Madrileña Foundation".
\end{acks}

\bibliographystyle{ACM-Reference-Format}
\balance
\bibliography{sample-base}

\newpage

\section*{Appendix A. Early vs. Late Fusion}
Table~\ref{tab:cm_late_detailed} (model I), which corresponds to the results for CMA as a late fusion strategy; and Table~\ref{tab:cm_unique_feature_detailed}, which contains the results for CMA as early fusion (model II), performing cross-modal attention
directly on embeddings, followed by a feedforward layer. Both tables contain extensive runs on the different modality combinations and key-query pairs. 

\begin{table}[h]
\centering
\small
\setlength{\tabcolsep}{4pt}
\caption{Detailed multi-modal results for the late fusion, using CMA as a fusion strategy. The features are concatenated if more than one modality is used for a key and value. Key~--~K:~key, Q:~query, T:~transcript, O:~on-screen text, A:~audio, V:~video, M:~macro average across both classes, H:~hate class, F1:~F1-score, ACC:~unbiased accuracy, P:~precision, R:~recall.}
\begin{tabular}{l|llccccc}
\hline
\textbf{Modality} & \textbf{K} & \textbf{Q} & \textbf{ACC} & \textbf{M-F1} & \textbf{F1(H)} & \textbf{P(H)} & \textbf{R(H)} \\
\hline
\multirow{2}{*}{TO}  & O  & T   & 0.658 & 0.630 & 0.527 & 0.617 & 0.460 \\
                     & T  & O   & \textbf{0.829} & \textbf{0.825} & \textbf{0.800} & \textbf{0.776} & \textbf{0.825} \\
\hline
\multirow{2}{*}{TA}  & A  & T   & \textbf{0.816} & 0.808 & 0.770 & \textbf{0.797} & 0.746 \\
                     & T  & A   & \textbf{0.816} & \textbf{0.811} & \textbf{0.781} & 0.769 & \textbf{0.794} \\
\hline
\multirow{2}{*}{TV}  & V  & T   & 0.737 & 0.733 & 0.701 & 0.662 & 0.746 \\
                     & T  & V   & \textbf{0.829} & \textbf{0.825} & \textbf{0.797} & \textbf{0.785} & \textbf{0.810} \\
\hline
\multirow{2}{*}{OA}  & A  & O   & 0.816 & 0.812 & \textbf{0.785} & 0.761 & \textbf{0.810} \\
                     & O  & A   & \textbf{0.829} & \textbf{0.820} & 0.780 & \textbf{0.836} & 0.730 \\
\hline
\multirow{2}{*}{OV}  & V  & O   & 0.743 & 0.735 & 0.688 & 0.694 & \textbf{0.683} \\
                     & O  & V   & \textbf{0.763} & \textbf{0.752} & \textbf{0.700} & \textbf{0.737} & 0.667 \\
\hline
\multirow{2}{*}{AV}  & V  & A   & 0.816 & 0.807 & 0.767 & 0.807 & \textbf{0.730} \\
                     & A  & V   & \textbf{0.822} & \textbf{0.812} & \textbf{0.769} & \textbf{0.833} & 0.714 \\
\hline
\multirow{3}{*}{TOA} & OA & T   & 0.796 & 0.791 & 0.760 & 0.742 & 0.778 \\
                     & TA & O   & \textbf{0.829} & \textbf{0.824} & \textbf{0.794} & \textbf{0.794} & 0.794 \\
                     & TO & A   & 0.809 & 0.807 & 0.785 & 0.736 & \textbf{0.841} \\
\hline
\multirow{3}{*}{TOV} & OV & T   & 0.770 & 0.759 & 0.711 & 0.741 & 0.683 \\
                     & TV & O   & \textbf{0.855} & \textbf{0.852} & \textbf{0.828} & \textbf{0.815} & \textbf{0.841} \\
                     & TO & V   & 0.829 & 0.824 & 0.797 & 0.784 & 0.809 \\
\hline
\multirow{3}{*}{TVA} & VA & T   & 0.796 & 0.785 & 0.735 & 0.796 & 0.682 \\
                     & TV & A   & 0.835 & 0.829 & 0.797 & \textbf{0.817} & 0.778 \\
                     & TA & V   & \textbf{0.842} & \textbf{0.839} & \textbf{0.818} & 0.783 & \textbf{0.857} \\
\hline
\multirow{3}{*}{OAV} & AV & O   & 0.796 & 0.789 & 0.752 & 0.758 & 0.746 \\
                     & OV & A   & \textbf{0.809} & 0.803 & 0.768 & \textbf{0.774} & 0.762 \\
                     & OA & V   & \textbf{0.809} & \textbf{0.805} & \textbf{0.775} & 0.757 & \textbf{0.794} \\
\hline
\multirow{4}{*}{TOAV} & OAV & T   & 0.789 & 0.786 & 0.758 & 0.725 & 0.794 \\
                      & TAV & O   & \textbf{0.882} & 0.877 & 0.852 & \textbf{0.881} & 0.825 \\
                      & TOV & A   & \textbf{0.882} & \textbf{0.879} & \textbf{0.862} & 0.836 & \textbf{0.889} \\
                      & TOA & V   & 0.842 & 0.837 & 0.810 & 0.810 & 0.810 \\
\hline
\end{tabular}
\label{tab:cm_late_detailed}
\end{table}

\vspace{200pt}
\begin{table}[h]
\centering
\small
\setlength{\tabcolsep}{4pt}
\caption{Detailed results for experiments using CMA as a fusion strategy on raw inputs. Key~--~K:~key, Q:~query, T:~transcript, O:~on-screen text, A:~audio, V:~video, M:~macro average across both classes, H:~hate class, F1:~F1-score, ACC:~unbiased accuracy, P:~precision, R:~recall.}
\begin{tabular}{l|llccccc}
\hline
\textbf{Modality} & \textbf{K} & \textbf{Q} & \textbf{ACC} & \textbf{M-F1} & \textbf{F1(H)} & \textbf{P(H)} & \textbf{R(H)} \\
\hline
\multirow{2}{*}{TO}  & O  & T   & 0.645 & 0.644 & 0.630 & 0.554 & 0.730 \\
                     & T  & O   & \textbf{0.829} & \textbf{0.825} & \textbf{0.800} & \textbf{0.776} & \textbf{0.825} \\
\hline
\multirow{2}{*}{TA}  & A  & T   & 0.809 & 0.806 & 0.779 & 0.750 & 0.809 \\
                     & T  & A   & \textbf{0.829} & \textbf{0.825} & \textbf{0.797} & \textbf{0.785} & \textbf{0.810} \\
\hline
\multirow{2}{*}{TV}  & V  & T   & 0.743 & 0.727 & 0.742 & 0.650 & 0.749 \\
                     & T  & V   & \textbf{0.836} & \textbf{0.832} & \textbf{0.809} & \textbf{0.779} & \textbf{0.841} \\
\hline
\multirow{2}{*}{OA}  & A  & O   & \textbf{0.783} & \textbf{0.776} & \textbf{0.736} & \textbf{0.742} & \textbf{0.730} \\
                     & O  & A   & 0.632 & 0.631 & 0.622 & 0.541 & \textbf{0.730} \\
\hline
\multirow{2}{*}{OV}  & V  & O   & \textbf{0.783} & \textbf{0.777} & \textbf{0.740} & \textbf{0.746} & \textbf{0.734} \\
                     & O  & V   & 0.651 & 0.651 & 0.634 & 0.561 & 0.730 \\
\hline
\multirow{2}{*}{AV}  & V  & A   & 0.770 & 0.760 & 0.711 & 0.741 & 0.683 \\
                     & A  & V   & \textbf{0.796} & \textbf{0.789} & \textbf{0.752} & \textbf{0.758} & \textbf{0.746} \\
\hline
\multirow{3}{*}{TOA} & OA & T   & 0.783 & 0.781 & 0.759 & 0.703 & 0.825 \\
                     & TA & O   & \textbf{0.816} & 0.811 & 0.781 & \textbf{0.769} & 0.793 \\
                     & TO & A   & \textbf{0.816} & \textbf{0.813} & \textbf{0.791} & 0.746 & \textbf{0.841} \\
\hline
\multirow{3}{*}{TOV} & OV & T   & 0.776 & 0.770 & 0.730 & 0.730 & 0.730 \\
                     & TV & O   & \textbf{0.842} & \textbf{0.839} & \textbf{0.818} & \textbf{0.783} & \textbf{0.857} \\
                     & TO & V   & 0.829 & 0.825 & 0.800 & 0.776 & 0.825 \\
\hline
\multirow{3}{*}{TVA} & VA & T   & 0.810 & 0.807 & 0.785 & 0.801 & 0.841 \\
                     & TV & A   & 0.862 & 0.858 & 0.835 & 0.82813 & 0.841 \\
                     & TA & V   & 0.816 & 0.812 & 0.785 & 0.761 & 0.810 \\
\hline
\multirow{3}{*}{OAV} & AV & O   & 0.770 & 0.769 & 0.752 & 0.679 & 0.841 \\
                     & OV & A   & 0.789 & 0.786 & 0.761 & 0.718 & 0.810 \\
                     & OA & V   & \textbf{0.809} & \textbf{0.807} & \textbf{0.788} & \textbf{0.730} & \textbf{0.857} \\
\hline
\multirow{4}{*}{TOAV} & OAV & T   & 0.822 & 0.820 & 0.797 & 0.757 & 0.841 \\
                      & TAV & O   & \textbf{0.888} & \textbf{0.884} & \textbf{0.864} & \textbf{0.871} & \textbf{0.860} \\
                      & TOV & A   & 0.855 & 0.851 & 0.825 & 0.851 & 0.825 \\
                      & TOA & V   & 0.842 & 0.840 & 0.824 & 0.767 & 0.889 \\
\hline
\end{tabular}
\label{tab:cm_unique_feature_detailed}
\end{table}

\newpage

\section*{Appendix B. Stopwords}

When removing stopwords from the transcript modality, we observe a small performance drop. The MM-HSD model with stopwords achieves an M-F1 of 0.874 and an F1-H of 0.853. After filtering out stopwords, M-F1 falls to 0.862, and F1-H to 0.841. This indicates that the model relies to some extent on certain high-frequency words from the transcript modality, such as negations, to identify HS. The full results are shown in Table~\ref{tab:stopwords_ablation}.

\begin{table}[h]
\centering
\small
\setlength{\tabcolsep}{1pt}
\caption{Effect of applying stopword removal to transcript and OCR modalities in MM-HSD. Key~--~M:~macro average across both classes, H:~hate class, F1:~F1-score, ACC:~unbiased accuracy, P:~precision, R:~recall.}
\begin{tabular}{l|ccccc}
\hline
\textbf{Model} & \textbf{ACC} & \textbf{M-F1} & \textbf{F1(H)} & \textbf{P(H)} & \textbf{R(H)} \\
\hline
\textbf{MM-HSD} & \textbf{.878 (.009)} & \textbf{.874 (.009)} & \textbf{.853 (.009)} & \textbf{.849 (.017)} & \textbf{.857 (.000)} \\
\makecell[l]{MM-HSD \\ (removing \\ stopwords)} & .866 (.006) & .862 (.006) & .841 (.006) & .826 (.011) & \textbf{.857 (.000)} \\
\hline
\end{tabular}
\label{tab:stopwords_ablation}
\end{table}

\section*{Appendix C. Analysis of Modalities Included in CMA}

Keeping the OCR as query, we experiment with removing modalities from the keys, while keeping all the modalities in the late fusion stage, with results presented in Table~\ref{tab:modality_combinations}. We observe that MM-HSD achieves the best performance across all metrics, suggesting that different modalities add complementary information. The worst modality is text-only, possibly due to the fact that there is some informational overlap between transcript and OCR, as on-screen text is sometimes subtitles from the audio track. However, audio-only is the second-best model, suggesting that audio contains different information than OCR, such as volume and emotion.

\begin{table}[h]
\centering
\small
\setlength{\tabcolsep}{4pt}
\caption{Performance of using excluding modalities from CMA. OCR is kept as the query modality in all cases.  
Key~--~M:~macro average across both classes, H:~hate class, \mbox{F1:~F1-score}.}
\begin{tabular}{l|cc}
\hline
\textbf{Modality} & \textbf{M-F1} & \textbf{F1(H)} \\
\hline
Audio only     & 0.870 & 0.848 \\
Video only     & 0.864 & 0.841 \\
Transcript only      & 0.855 & 0.832 \\
Audio + Video  & 0.866 & 0.845 \\
Audio + Transcript   & 0.859 & 0.834 \\
Video + Transcript   & 0.861 & 0.837 \\
\textbf{MM-HSD (A+V+T)} & \textbf{0.874} & \textbf{0.853} \\
\hline
\end{tabular}
\label{tab:modality_combinations}
\end{table}

\vspace{400pt}
\end{document}